# Health Wars and Beyond:
# The Rapidly Expanding and Efficient Network Insurgency Interlinking Local and Global Online Crowds of Distrust

N.F. Johnson[1,2], N. Velasquez[2], N. Johnson Restrepo[1,2], R. Leahy[1,2], N. Gabriel[1], S. Wuchty[3], D.A. Broniatowski[4,2]
[1]*Physics Department, George Washington University, Washington D.C. 20052*
[2]*Institute for Data, Democracy and Politics, George Washington University, Washington D.C. 20052*
[3]*Department of Computer Science, University of Miami, Coral Gables FL 33124*
[4]*School of Engineering and Applied Sciences, George Washington University, Washington D.C. 20052*

**We present preliminary results on the online war surrounding distrust of expertise in medical science – specifically, the issue of vaccinations. While distrust and misinformation in politics can damage democratic elections, in the medical context it may also endanger lives through missed vaccinations and DIY cancer cures. We find that this online health war has evolved into a highly efficient network insurgency with direct inter-crowd links across countries, continents and cultures. The online anti-vax crowds (referred to as 'Red') now appear better positioned to groom new recruits ('Green') than those supporting established expertise ('Blue'). We also present preliminary results from a mathematically-grounded, crowd-based analysis of the war's evolution, which offers an explanation for how Red seems to be turning the tide on Blue.**

Distrust and misinformation pose an acute global threat to established science and medicine, as well as political processes [1-10]. Death threats are being made against climate scientists [3], and doctors and mothers who vaccinate [4]. In addition to widespread spreading of diseases such as measles and HPV through vaccination hesitancy [5-8], distrust of established scientific treatments for cancer are leading patients to adopt dangerous substitute measures [9]. More generally, the number of countries with disinformation campaigns is reported to have more than doubled to 70 in the last two years, with Facebook remaining the top platform for those campaigns [10]. Given the difficulty social media companies such as Facebook are having with online activity related to hate [11], terrorism [12] and child sexual abuse [13,14], it is no wonder that they are also struggling to understand how such online misinformation develops and spreads from local to global scales, and hence how to go about dealing with it. There have been many valuable studies of distrust, misinformation and disinformation at the level of individuals' behaviors including on Twitter. However, DiResta and others have pointed out that what is missing is a big picture understanding in the national and global populations involving millions of people [15] while Starbird has also highlighted the need to understand the important role of unwitting crowds [16,17]. Indeed, the impact of crowds -- and in particular undecideds -- is known to play a central role in the dynamics of other societal systems such as financial markets [17] and elections [1,2,18].

This short note provides some preliminary findings from our system-level analysis of the current health war over vaccinations in the universe of ~3 billion social media users worldwide, focusing on these online crowds. There are of course caveats to our study and analysis, which we discuss later in this document. However, while these may affect the details of what we show, we see no reason why they should qualitatively distort our main findings and hence we believe that our main conclusions are robust. We will be carrying out further tests and collecting more data in the near future. When eventually in final form, the figures presented here will look different, not least since the data will be updated, but we expect that the main takeaways will remain similar.

Figure 1 shows the largest component in the network of >10 million people who – as of Oct 1, 2019 – have decided to partition themselves into clusters (e.g. Facebook Pages) that are either distrustful of established medical science surrounding vaccinations ('Red', anti-vaccination), or in support it ('Blue', pro-vaccination), or as yet undecided but nonetheless engaged ('Green', neutral). A Green cluster is considered to be engaged in some way if this cluster is connected to another Red or Blue cluster in the network by a direct link at the cluster level -- not simply that they have a few followers in common with each other. This method of focusing on direct links between clusters at the cluster level, is the same as in our previously published studies of online pro-ISIS support [12] and online hate [11]. Our focus on clusters – and in particular, the mesoscale cluster dynamics of Red, Blue and Green -- is consistent with the fact that clustered correlations are known to provide the key to understanding the dynamics of many-body physical, chemical and biological systems and the likely extension to social systems. Our focus on clusters also conveniently bypasses the current data impasse facing researchers, and facing Facebook itself, over concerns for individual privacy.

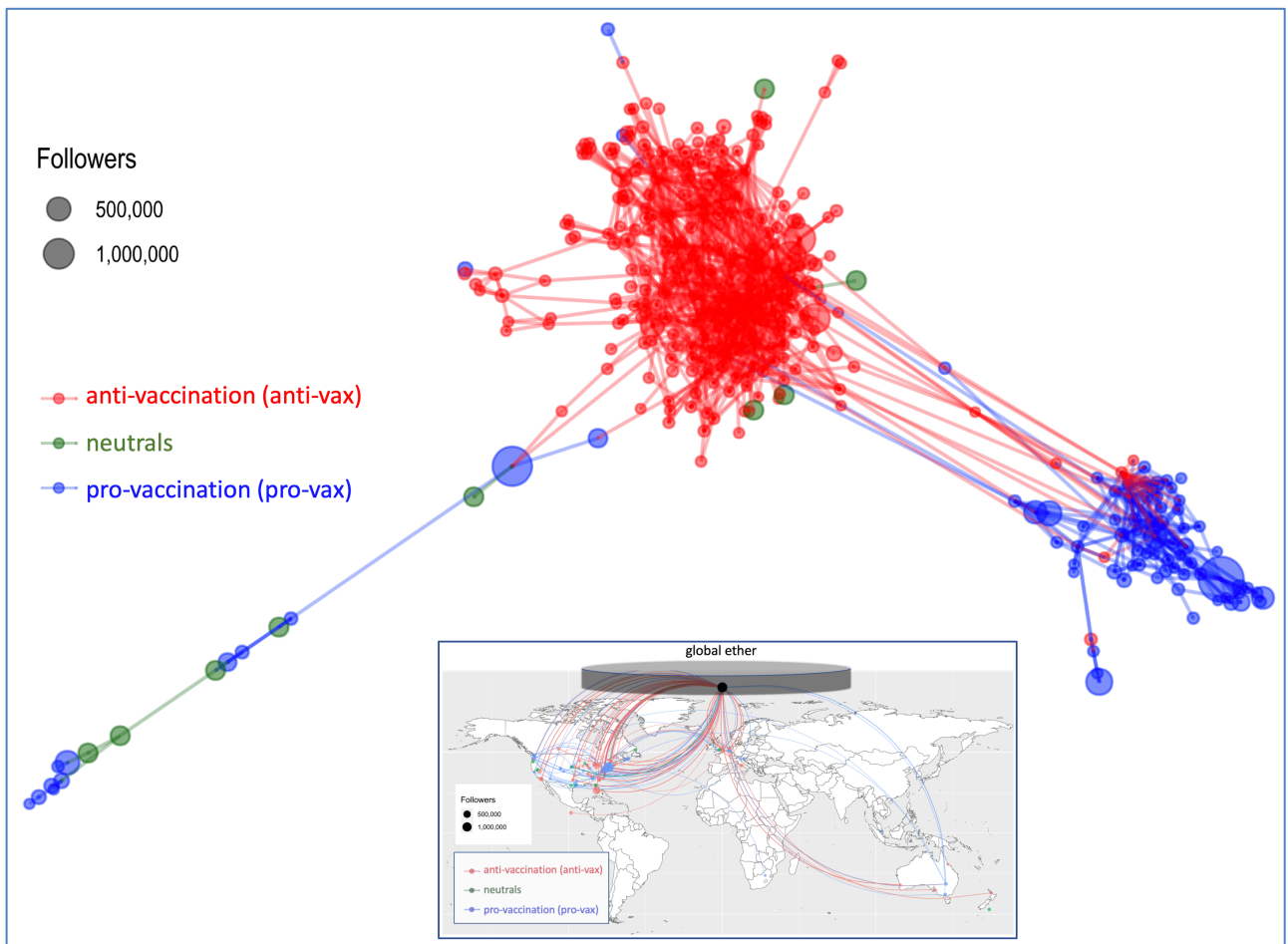

**Figure 1. Health war raging online globally surrounding vaccines.** This is the largest component of the network. It is drawn from tracking the online clusters (e.g. Facebook pages) that are anti-vaccination (Red), pro-vaccination (Blue) and neutral (Green). Inset shows a map with these clusters according to their declared geographical location (see Fig. 2 for more details). These results, while preliminary due to missing data about other Green clusters and current cross-checking of Red and Blue, already show the tendency for the Red subpopulation to form into a large number of clusters with a range of sizes, that are connected together in a seemingly decentralized way – akin to an insurgency. Red clusters are more closely engaged with Green clusters than Blue clusters are. The Blue clusters tends to be off to one side of the online war, with large formal organizations such as the Gates Foundation (large blue circle, bottom right) seemingly set back from this battlefield. Cluster network drawn using ForceAtlas2 algorithm. A full version of the network (i.e. with a full search for connected Greens) is given in Fig. 4, and shows similar features, with the trivial presentation change that the side-arm of Blue then appears on the right.

There are several features of Fig. 1 that make this online health war bear more than a passing analogy to an insurgency. The Green clusters represent undecided, or swing, or unwitting crowds, and the battle for their support is akin to the battle for the 'hearts and minds' of a local civilian population that becomes the focus in insurgent warfare. Red forms into a large number of clusters (approx. 490) with sizes ranging from less than 100 to several hundreds of thousands. Moreover, they are connected together in a seemingly decentralized way, akin to an insurgency [19]. In addition, Blue is numerically larger (nearly 7 million) compared to Red (approximately 4 million) and yet there are many more clusters of Red. Instead Blue is organized into large clusters akin to a conventional army (e.g. Gates Foundation cluster size is nearly 1.5 million). Though Blue is numerically larger than Red, it can be seen from Fig. 1 that Red and Green are overall far more heavily engaged with each other than the case for Blue and Green. Again, this is typical of insurgencies where insurgents are usually more successful in engaging the local civilian (i.e. as-yet non-combatant) population than Blue. It is also telling that each Red cluster has on average more links to other Red clusters than each Blue cluster has on average with other Blue clusters, suggesting that Red looks to build out connections to other Red while Blue tends to have more stand-alone clusters. Moving to the dynamics: though both Red and Blue clusters change size between June-October 2019, some Red insurgent clusters grew more than 300% since early summer 2019 with one exceeding 500%. In contrast, the maximum growth of any Blue cluster was less than 100%. Such a rapid rise in the size of small clusters is typical of an insurgency – and in particular, it is exactly what we observed previously for online support of a real insurgency in Ref. 12.

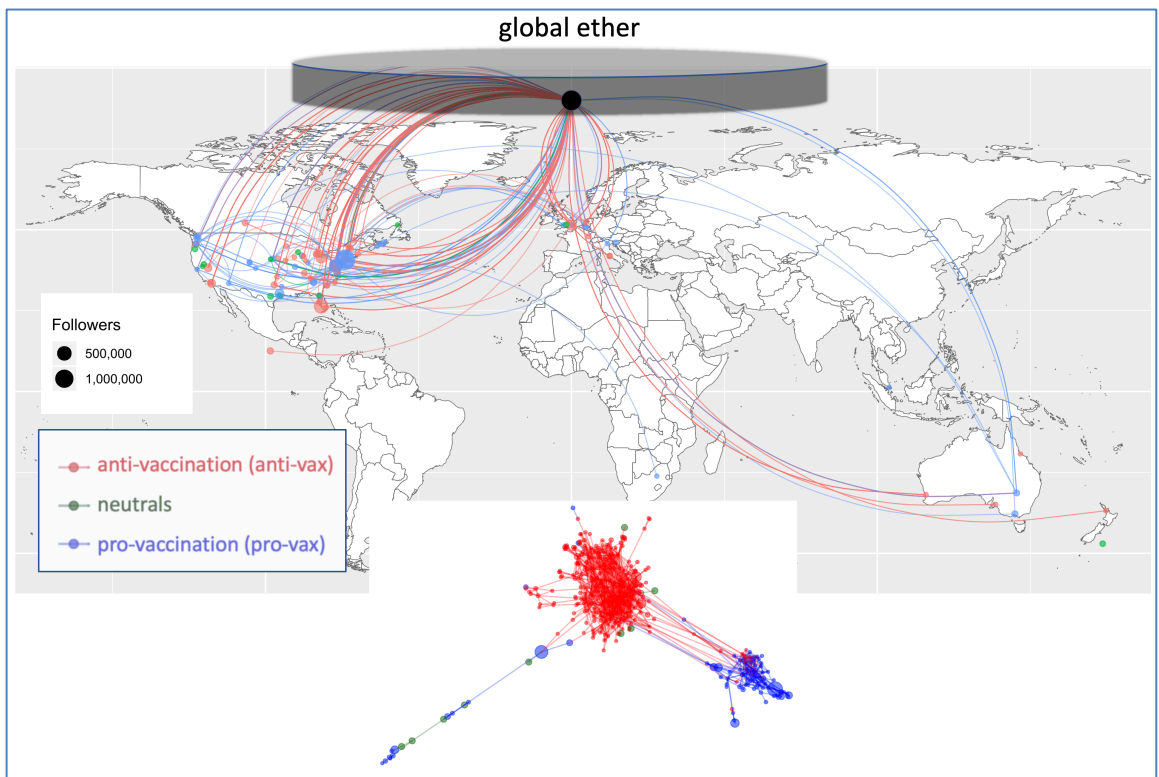

**Figure 2. Health war showing clusters from the network in Fig. 1 mapped onto their declared geographical location. Inset shows the cluster network drawn without regard to location, and instead with an arrangement based on closeness of links using ForceAtlas2 algorithm (see also Fig. 1). These results, while again preliminary due to missing data about other Green clusters and the need to cross-check Red and Blue, show how clusters connect this online war across countries and continents. The "global ether" is made up of those clusters who choose not to localize themselves further than the global scale – and hence which presumably see themselves as a global cause or community.**

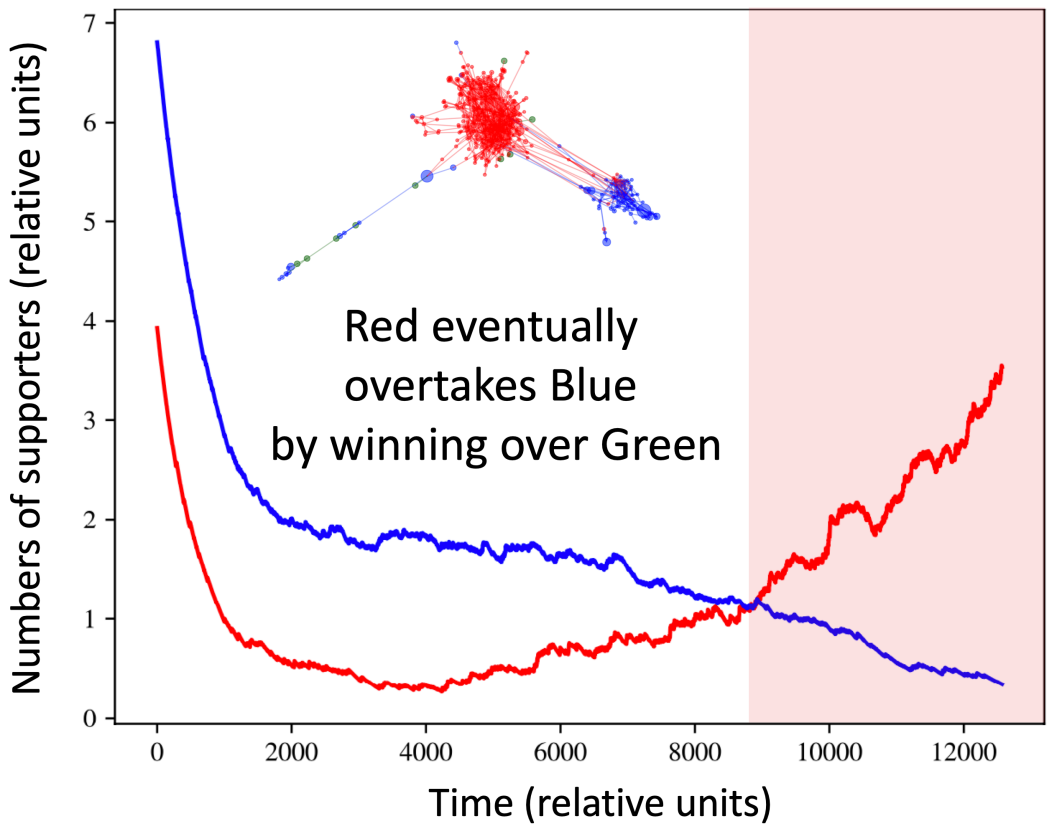

**Figure 3. Preliminary results from our mathematically-grounded, crowd-based analysis of the war's evolution, which offers an explanation for how Red turns the tide against Blue. A full version of the network (i.e. with a full search for connected Greens) is given in Fig. 4, and shows similar features.**

Figure 3 shows the output of our accompanying model. Leaving mathematical details for a longer paper, we note here that the model comprises an ecology of clusters (Red, Blue and Green) with probabilities of interactions in time given by the relative number of links of each type in Fig. 1. The rules of engagement are that Red is, when interacting with Green, effective in being able to turn Green over to its side, while Blue is not. We also add the feature that Red and Blue have a natural loss of interest over time, and that Red and Blue can interact and reduce each others' numbers to mimic people getting tired of the fight or realizing the other side's viewpoint. Of course, these rules can be changed and nuanced, and need to be explore more deeply – but it is already noteworthy that, as shown in Fig. 3, Red starts from a minority position but is able to eventually show an upsurge that makes them dominant over Blue.

Figure 4 shows a fuller version of the network from Fig. 1, obtained by carrying out a full search for connected Greens. It shows similar features to Fig. 1, but serves to show more explicitly how Red clusters are heavily embedded among many Green clusters while Blue clusters are not. As a result of this expanded search for Greens, there are now approximately 100 million individuals in Fig. 4.

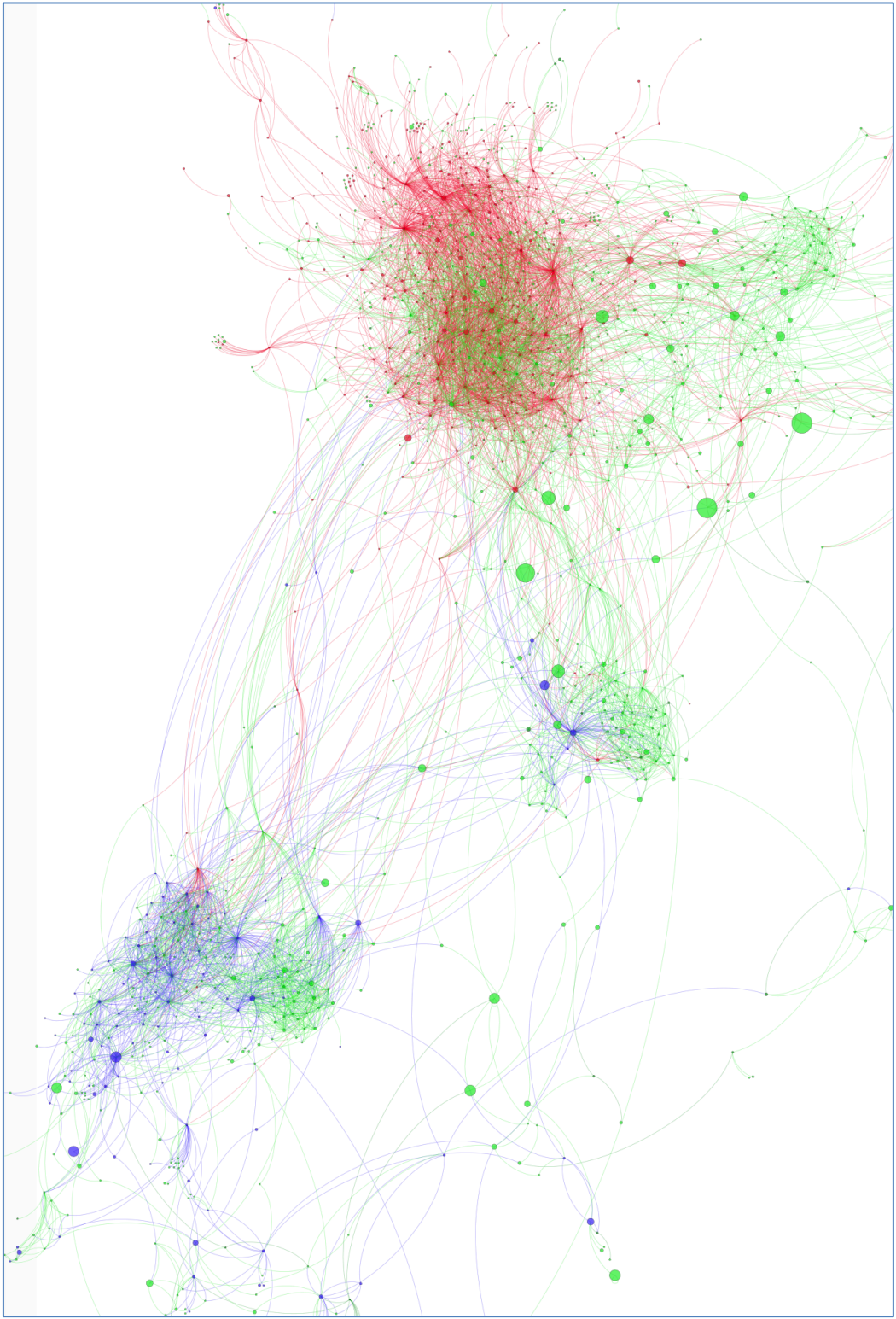

**Figure 4: Fuller version of Fig. 1's health war network, obtained using a more complete search for connected Greens. It shows similar features to Fig. 1, but serves to show more explicitly how Red (anti-vax) clusters are heavily embedded among Green (neutral) clusters while Blue (pro-vax) clusters are not. There is a trivial presentation change compared to Fig. 1: it is rotated and flipped with the side-arm of Blue now appearing on the right. There are approximately 100 million individuals included in this plot.**

There are of course limitations of our study. We have only presented results for one type of distrust – however our preliminary findings for political and commercial scenarios are somewhat similar (i.e. fans of product X versus fans of competing product Y). Also, we have not touched the details of the social network in Figs. 1 or 2. Instead our focus here is on the broad-brush system level behavior. We also need to look across other social media platforms, but it is likely to be similar on any platform that allows communities to be built and hence where deeper debate can be developed. Also, our quantitative analysis is highly idealized in order to generate quantitative answers. Although not intended to capture the complications of any specific real- world setting, the benefit of the modelling approach in Fig. 3 is that the output is precisely quantified, reproducible and generalizable, and can therefore help to frame policy discussions as well as probe what-if intervention scenarios. One may also wonder about the role of external agents or entities, including not only bots but State actors. In terms of bots, it tends to be that clusters police themselves for any postings with robotic formats to weed out such infiltrations. In terms of State actors, they may well be present and our analysis in this paper neither supports nor denies their presence. Despite these and other potential critiques, we believe that an understanding of how undecideds become drawn into the distrust ecologies online, may also help with other currently 'lost' online wars where the core narratives and intent to act are discussed and developed in online clusters. One example is online child sexual abuse which Refs. 13,14 describe as at a breaking point, with reports of abusive images exceeding the capabilities of independent clearinghouses and law enforcement to take action. Another potential application is to politics and the upcoming 2020 elections in the U.S.